\documentstyle[psfig]{l-aa}

\begin{document}

\thesaurus{03.06.1; Pleiades; 18.02.1; 19.71.1}
\title{Investigation of the Pleiades cluster\thanks{based on observations
collected at the Haute-Provence Observatory (France). Table 3 is
available only in electronic form from the Strasbourg ftp server at
130.79.128.5}}

\subtitle{IV. The radial structure}
\author{D.~Raboud \and J.-C.~Mermilliod}


\institute{Institut d'Astronomie de l'Universit\'{e} de Lausanne, CH-1290 
Chavannes-des-Bois}

\date{Received date; accepted date}

\maketitle

\begin{abstract}
On the basis of the best available member list and duplicity information, we 
have studied the radial distribution of 270 stars and multiple systems earlier than K0 in the Pleiades. Five new long period spectroscopic binaries have been identified from the CORAVEL observations. We have found a clear mass segregation between binaries and single stars, which is explained by the greater average mass of the multiple systems. The mass function of the single stars and primaries appears to be significantly different. While the central part of the cluster is spherical, the outer part is clearly elliptical, with an ellipticity of 0.17. 
The various parameters describing the Pleiades are (for a distance of 125 pc): 
core radius $r_{c} = 0\fdg6$ (1.4 pc), tidal radius $r_{t} = 7\fdg4$ (16 pc), 
half mass radius $r_{m/2} = 0\fdg88$ (1.9 pc), harmonic radius 
$\overline{r} = 1\fdg82$ (4 pc). 
Low-mass stars (later than K0) probably extend further out and new proper motion 
and radial velocity surveys over a larger area and to fainter magnitudes would 
be very important to improve the description of the cluster structure and
complete mass function.

\keywords{Clusters: open - Individual: Pleiades - Structure - technique:
radial velocity - binary: spectroscopic}

\end{abstract}

\section{Introduction}
The unity of star cluster structures has been discussed about thirty years
ago by Kholopov (1969) as a generalization of the results obtained first
on the basis of star counts, and later of proper motion studies, covering
wide areas in several nearby open clusters made by Artyukhina in the Pleiades
(Artyukhina 1969; Artyukhina \& Kalinina 1970), Praesepe (Artyukhina 1966) and other nearby clusters. He clearly showed the existence of a core - halo
structure and got a tentative relation between the core and tidal radii.
The star count methods give only statistical results because no real knowledge 
of the member stars is obtained. Results based on proper motion studies
are more sound, because cluster members are individually selected. 

The core - halo structure of open clusters based on the direct identification 
of individual member stars has so far been established for a small number 
of open clusters, among them Alpha Per (Artyukhina 1972), the Hyades (Oort 1979), NGC 6705 (Mathieu 1984; Solomon \& McNamara 1980), NGC 2682 (Mathieu \& Latham 1986). This is easily explained by the difficulty of identifying the true members in the outer part of the clusters where the density of cluster members is low with respect to the projected density of field stars, i.e. one star per square degree at 2$\fdg$5 from the center in Praesepe. Due to the considerable area occupied over the sky by nearby clusters with diameters of about 10${\degr}$, the number of stars to measure for proper motions reaches several times 10$^4$. However, a definitive understanding of the cluster structure and dynamical evolution can only be obtained through a detailed knowledge of the properties of the member stars. Pleiades and Praesepe are, among the nearby 
clusters, primary candidates to study their overall structure, because of the 
possibility of obtaining very detailed information on individual stars.

Mass segregation and concentration of binary stars towards the cluster
centers are predicted by theories and numerical simulations of cluster dynamical evolution (Spitzer \& Mathieu 1980, Kroupa 1995, de la Fuente Marcos 1996), but observational evidences have so far also been difficult to gather because of the generally too limited cluster surface coverage and the lack of systematic radial velocity survey or detection of binaries in the range of 0\farcs05 to 1\farcs0. This traditional gap in binary period coverage is now beginning
to be filled by speckle interferometry observations obtained at CHARA
(Mason et al. 1993) and direct imaging in the near-IR with adaptive 
optics (Bouvier et al. 1997). But it is still true that radial
velocity surveys of the upper main sequences in the Pleiades, Praesepe
and $\alpha$ Persei clusters are badly needed to improve the knowledge
of binarity and orbit characteristics to the level reached by the
surveys of the solar-type stars (Duquennoy et al. 1992).

Previous studies of the structure of the Pleiades (Artyukhina 1969; Peikov 
1990) relied on proper motions to detect the member stars in the outer
part of the cluster, but without photometric data to enable an analysis 
of the membership in the colour-magnitude diagram. van Leeuwen's very comprehensive study was based on photometric observation in the 
Walraven photometric system and independent new proper motions (van Leeuwen 1983, van Leeuwen et al. 1986). However, the studies of the coronas of the Pleiades (Rosvick et al. 1992a) and of Praesepe (Mermilliod et al. 1990) showed that accurate proper motions, reliable photometry and radial velocities are needed to determine reliable membership estimates for the stars in the very outer part of the clusters.

New data are now available for the Pleiades due to the systematic radial 
velocity survey undertaken by one of the authors (JCM) with the CORAVEL scanner 
on the lower main sequence, the speckle interferometry survey done at CHARA (Mason et al. 1993; Mason 1996) and the near-IR imaging (Bouvier et al. 1997). This unique material describes for the first time the binary characteristics
of the cluster stars and offers the opportunity of studying the radial structure and mass distribution in the Pleiades with much more detail.

The first and third papers in this series on the Pleiades (Rosvick et al. 1992a; Mermilliod et al. 1997) have specified the membership of stars in the corona of this cluster. In addition to proper motions and photometry from the literature, we have used radial velocity to improve the selection. We have shown that about 50\% of the cluster members in the spectral range F5-K0 are located outside the classical area studied by Hertzsprung (1947) and member stars have been found as far as nearly 5$\degr$ from the center. This
makes the real size of the cluster much larger. The improved list of cluster members bears a lot of importance in the determination of the mass function of the Pleiades. Searches for lower mass stars, extending those conducted in the central part (Stauffer et al. 1991, Jameson 1993), will certainly reduce the discrepancies between the field star mass function and those observed in star clusters, usually explained by cluster evaporation.

This paper will discuss the new mass function and the radial structure of 
the Pleiades, with emphasise on mass segregation for the brighter and fainter 
stars, or between the single and binary members.
We first present the results of the radial velocity survey for the
central part of the Pleiades and the orbit of a new binary (Sect. 2),
discuss the binarity of the upper main sequence (Sect. 3), describe the 
catalogue of member stars (Sect. 4) and present the results on the cluster 
structure and the mass function in Sect. 5.

\section{Observations and Results}
\subsection{CORAVEL observations}
Previous papers presented the radial velocities obtained for stars in the 
corona (Rosvick et al. 1992a; Mermilliod et al. 1997), but so far the data 
for stars in Hertzsprung (1947) catalogue have not been published.
The observations were obtained with the CORAVEL radial-velocity scanner 
(Baranne et al. 1979) installed on the Swiss 1-m telescope at the 
Haute-Provence Observatory (OHP) for stars later than spectral type F5 and
brighter than $B$ = 12.5 - 13.0. A few stars, mainly F5 stars, 
could not be observed because of their too large rotation. 
Between September 1978 and December 1995, five to nine observations per star 
were obtained. It has not been possible to get recent epoch radial velocity observations for the complete sample, but a large fraction has been reobserved to improve the detection of longer period binaries. Indeed, most orbits determined so far in the Pleiades are shorter than 1000 days (Mermilliod et al. 1992). One more orbit has been determined here for HII 1338, with a period close to 8 days. Five new spectroscopic binaries have been detected (HII 120, 263, 476, 916
and 2500) with velocity ranges of a few km s$^{-1}$ (3.1, 4.5, 4.6, 4.1, 8.9).
The radial velocities of HII 263 and 476 show a clear drift over 15 years
of observations. Double lines have been observed for HII 1122 by Liu et al. (1991) on JD 2447128.7, with velocities of 41.8 and -79.1 km s$^{-1}$ for components A and B respectively. Our observations show only one dip.

The HII numbers (Hertzsprung 1947), the $V$ magnitudes, the mean radial velocities $V_{r}$, the standard errors $\epsilon$ (in km s$^{-1}$), the number of measurements $n$, the time intervals $\Delta T$ covered by the observations, the probability $P(\chi^2)$ that the scatter is due to chance (Mermilliod \& Mayor 1989) are collected in Table 1. The remarks SB1O or SB2O refer to orbits obtained by Mermilliod et al. (1992), PHB to binaries detected in the colour-magnitude diagram from the same reference, IRB to binary found by infra-red imaging (Bouvier et al. 1997), VB to visual binary. Individual data are available upon request (Jean-Claude.Mermilliod@obs.unige.ch). The whole CORAVEL dataset is currently being recalibrated for possible zero-point error and color effect. All the individual data will then be published in a comprehensive catalogue of observations in the Pleiades. 

\begin{table}
\caption[]{CORAVEL data for members of the Pleiades}
\label{mem}
\begin{flushleft}
\begin{tabular}{rrrrrrrl}
\hline\noalign{\smallskip}
\multicolumn{1}{c}{HII} & $m_{V}$ & $V_{r}$ & \multicolumn{1}{c}{$\epsilon$} & $n$ & $\Delta T$ & $P(\chi^2)$ & Remarks \\
\hline\noalign{\smallskip}
  25 &  9.47 &   4.9 &  0.9 &   5 &  4685 & 0.440 & \\
  34 & 11.99 &   4.9 &  0.2 &   4 &  3631 & 0.420 & \\
 102 & 10.51 &   3.9 &  0.2 &  19 &  5559 & 0.052 & IRB, 3$\farcs$63 \\
 120 & 10.79 &   5.7 &  0.6 &   5 &  4765 & 0.000 & SB1, new \\
 129 & 11.47 &   6.5 &  0.2 &   7 &  4765 & 0.053 & \\
 152 & 10.73 &   5.2 &  0.2 &   5 &  4765 & 0.411 & \\
 164 &  9.53 &   9.0 &  4.2 &   4 &  2124 & 0.000 & SB1 \\
 173 & 10.87 &   4.8 &  0.1 &  51 &  4406 & 0.000 & SB2O \\
 186 & 10.51 &   5.5 &  0.2 &   8 &  3624 & 0.271 & PHB \\
 193 & 11.29 &   7.6 &  0.2 &   7 &  4765 & 0.169 & \\
 233 &  9.66 &   5.6 &  0.6 &   8 &  5833 & 0.000 & SB1 \\
 248 & 11.02 &   6.3 &  0.4 &   5 &  4763 & 0.031 & \\
 250 & 10.70 &   4.7 &  0.5 &   6 &  4733 & 0.000 & SB ? \\
 253 & 10.66 &   4.2 &  0.8 &   6 &  4780 & 0.039 & \\
 263 & 11.54 &   4.6 &  0.5 &   9 &  5822 & 0.000 & SB, new \\
 293 & 10.79 &   5.6 &  0.2 &   7 &  5020 & 0.610 & \\
 296 & 11.46 &   5.8 &  0.4 &   7 &  5526 & 0.000 & SB? \\
 298 & 10.86 &   4.8 &  0.2 &   6 &  4763 & 0.473 & IRB, 5$\farcs$69 \\
 303 & 10.48 &   6.0 &  0.5 &   7 &  5053 & 0.001 & VB, 1$\farcs$7 \\
 314 & 10.58 &   4.3 &  0.7 &  10 &  5020 & 0.008 & \\
 320 & 11.03 &   5.7 &  0.2 &  26 &  5559 & 0.000 & SB1O \\
 338 &  9.06 &   9.7 &  1.0 &   5 &  5049 & 0.719 & \\
 345 & 11.61 &   4.7 &  0.3 &   8 &  5557 & 0.072 & \\
 405 &  9.82 &   3.9 &  0.3 &   8 &  5021 & 0.351 & \\
 430 & 11.40 &   4.6 &  0.2 &   6 &  5526 & 0.566 & \\
 470 &  8.90 &   6.2 &  0.6 &   5 &  3614 & 0.591 & \\
 476 & 10.80 &   5.4 &  0.7 &   7 &  2958 & 0.000 & SB, new \\
 489 & 10.38 &   4.9 &  0.3 &   9 &  5107 & 0.099 & \\
 514 & 10.71 &   4.8 &  0.3 &   6 &  4818 & 0.091 & \\
 522 & 11.96 &   6.3 &  0.1 &  22 &  2925 & 0.000 & SB1O \\
 530 &  8.95 &   5.6 &  0.6 &   6 &  4376 & 0.011 & \\
 571 & 11.23 &   5.7 &  0.1 &  24 &  4404 & 0.000 & SB1O \\
 627 &  9.66 &   5.6 &  1.3 &   6 &  4685 & 0.000 & SB ? \\
 659 & 12.09 &   5.7 &  0.3 &   4 &  3309 & 0.596 & \\
 739 &  9.55 &   5.8 &  0.2 &  12 &  5561 & 0.016 & PHB \\
 746 & 11.28 &   6.5 &  0.3 &   9 &  5559 & 0.000 & SB ? \\
 761 & 10.56 &   5.4 &  0.1 &  33 &  4693 & 0.000 & SB1O \\
 879 & 12.79 &   4.6 &  0.4 &   3 &  1483 & 0.034 & \\
 885 & 12.06 &   5.2 &  0.3 &   4 &  4105 & 0.038 & IRB, 0$\farcs$87 \\
 916 & 11.71 &   5.4 &  0.8 &   5 &  5523 & 0.000 & SB, new \\
 923 & 10.14 &   6.4 &  0.3 &  11 &  5832 & 0.052 & \\
 975 & 10.57 &   4.9 &  0.4 &  26 &  5831 & 0.003 & PHB \\
 996 & 10.41 &   6.1 &  0.3 &   7 &  5020 & 0.156 & \\
1015 & 10.55 &   5.8 &  0.3 &   6 &  4731 & 0.190 & \\
1032 & 11.22 &   4.1 &  0.5 &   6 &  5522 & 0.549 & \\
1095 & 11.92 &   4.8 &  0.3 &   4 &  4105 & 0.067 & \\
1101 & 10.27 &   5.4 &  0.3 &   8 &  5020 & 0.463 & \\
1117 & 10.20 &   7.1 &  0.1 &  45 &  3756 & 0.000 & SB2O \\
1122 &  9.28 &   5.0 &  0.6 &   7 &  5750 & 0.252 & SB2 \\
1124 & 12.22 &   5.6 &  0.5 &   3 &  1773 & 0.052 & \\
1132 &  9.42 &   5.2 &  1.3 &   5 &  5055 & 0.047 & \\
1139 &  9.38 &   6.4 &  1.0 &   5 &  4685 & 0.020 & \\
1182 & 10.46 &   6.2 &  0.3 &   5 &  5526 & 0.956 & IRB, 1$\farcs$14\\
1200 &  9.90 &   5.7 &  0.5 &   5 &  4403 & 0.031 & \\
1207 & 10.47 &   5.5 &  0.2 &   6 &  4731 & 0.291 & \\
1215 & 10.53 &   6.0 &  0.2 &   7 &  5751 & 0.753 & \\
\noalign{\smallskip}
\hline
\end{tabular}
\end{flushleft}
\end{table} 

\setcounter{table}{0}
\begin{table}
\caption[]{(continued)}
\label{mem2}
\begin{flushleft}
\begin{tabular}{rrrrrrrl}
\hline\noalign{\smallskip}
\multicolumn{1}{c}{HII} & $m_{V}$ & $V_{r}$ & $\epsilon$ & $n$ & $\Delta T$ & $P(\chi^2)$ & Remarks \\
\hline\noalign{\smallskip}
1220 & 11.74 &   6.1 &  0.2 &   4 &  4732 & 0.503 & \\
1275 & 11.46 &   5.4 &  0.2 &   6 &  5526 & 0.398 & \\
1298 & 12.26 &   6.4 &  0.4 &   3 &  1506 & 0.079 & IRB, 1$\farcs$18 \\
1332 & 12.46 &   5.6 &  0.2 &   3 &  1483 & 0.874 & \\
1338 &  8.69 &   6.3 &  0.4 &  25 &   773 & 0.000 & SB2O \\
1392 &  9.50 &   6.7 &  0.5 &   4 &  1505 & 0.068 & \\
1514 & 10.48 &   4.9 &  0.3 &   7 &  5108 & 0.168 & \\
1593 & 11.11 &   6.9 &  0.2 &   6 &  5525 & 0.669 & \\
1613 &  9.87 &   5.7 &  0.3 &   7 &  4694 & 0.449 & \\
1726 &  9.27 &   5.8 &  0.3 &   8 &  5462 & 0.035 & VB, 0$\farcs$57 \\
1766 &  9.13 &   6.5 &  0.5 &   7 &  4269 & 0.141 & \\
1776 & 10.92 &   5.8 &  0.2 &   6 &  5522 & 0.727 & \\
1794 & 10.36 &   5.7 &  0.2 &   7 &  5108 & 0.397 & \\
1797 & 10.10 &   5.9 &  0.4 &   7 &  5751 & 0.066 & \\
1856 & 10.02 &   6.1 &  0.4 &   7 &  4694 & 0.029 & \\
1924 & 10.34 &   5.6 &  0.3 &   7 &  5108 & 0.266 & \\
2027 & 10.91 &   4.4 &  0.3 &  40 &  5881 & 0.000 & SB1O, IRB, 0$\farcs$1 \\
2106 & 11.53 &   4.9 &  0.2 &   6 &  5554 & 0.899 & IRB, 0$\farcs$32 \\
2126 & 11.68 &   5.5 &  0.2 &   6 &  5554 & 0.656 & \\
2147 & 10.85 &  11.1 &  1.0 &  36 &  5828 & 0.000 & SB2 \\
2172 & 10.44 &   5.9 &  0.1 &  34 &  4693 & 0.000 & SB1O \\
2278 & 10.91 &   4.5 &  0.2 &   9 &  5822 & 0.034 & IRB, 0$\farcs$37 \\
2284 & 11.35 &   6.3 &  1.1 &   8 &  5874 & 0.000 & SB1 \\
2311 & 11.35 &   5.4 &  0.2 &   5 &  5554 & 0.545 & \\
2341 & 10.87 &   6.7 &  0.2 &   5 &  4728 & 0.474 & \\
2366 & 11.53 &   6.2 &  0.2 &   6 &  5555 & 0.299 & \\
2406 & 11.10 &   6.0 &  0.1 &  25 &  4401 & 0.000 & SB1O \\
2407 & 12.24 &   5.7 &  0.1 &  21 &  2562 & 0.000 & SB1O \\
2462 & 11.52 &   6.4 &  0.2 &   6 &  5555 & 0.577 & \\
2500 & 10.20 &   5.7 &  1.4 &   6 &  1740 & 0.000 & SB, new \\
2506 & 10.23 &   6.1 &  0.4 &   7 &  4694 & 0.004 & \\
2644 & 11.05 &   4.9 &  0.2 &   5 &  5522 & 0.153 & \\
2665 & 11.36 &   7.1 &  0.2 &   5 &  5555 & 0.539 & \\
2786 & 10.31 &   5.2 &  0.4 &   6 &  5520 & 0.502 & \\
2880 & 11.75 &   5.4 &  0.3 &   6 &  5555 & 0.083 & \\
2881 & 11.55 &   5.2 &  0.2 &   7 &  5555 & 0.187 & IRB, 0$\farcs$08 \\
3096 & 12.12 &   5.8 &  0.2 &   3 &  1515 & 0.616 & \\
3097 & 10.97 &   5.6 &  0.2 &  22 &  5555 & 0.000 & SB1O \\
3179 & 10.04 &   6.1 &  0.2 &   5 &  4403 & 0.724 & \\
\noalign{\smallskip}
\hline
\end{tabular}
\end{flushleft}
\end{table} 

\subsection{The orbit of HII 1338}
\begin{figure}[t]
\centerline{\psfig{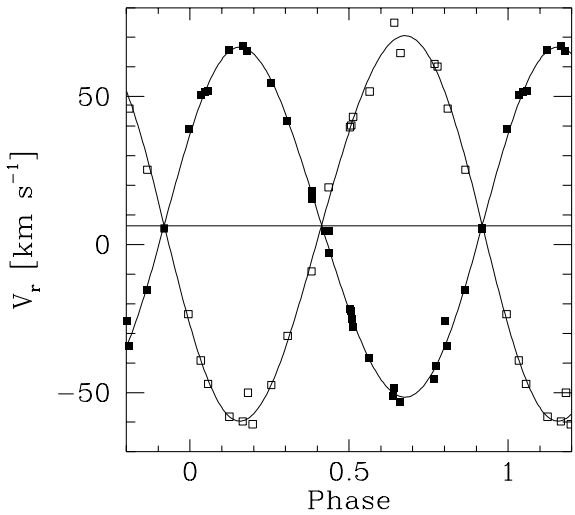}}
\caption[]{Radial-velocity curve for HII 1338}
\label{e1338}
\end{figure}

HII 1338 had been initially discarded because of the large published 
$V\sin i$ value, 110 km s$^{-1}$ (Kraft 1967), which was an 
artefact due to the double-lined character of this star discovered by 
Soderblom et al. (1993). We have therefore observed this star over two seasons and got a quasi-circular orbit (Fig.~\ref{e1338}), well defined with 25 and 21 observations of the primary and secondary stars respectively. This system proved to be interesting  because the period (Table~\ref{orb}) is 
very close to the theoretical cut-off period for circular orbit predicted 
by Zahn \& Bouchet (1989). The cross-correlation dips exhibit a rather small constrast, which explains the higher than usual value of the $O-C$ residuals. There is a hint of a third companion, which produces a dip visible when the primary components are at maximum separation in radial velocity and with a velocity close to the cluster value. Further observations will be done to clarify this point.

\begin{table}
\caption{Orbital elements of HII 1338}
\label{orb}
\begin{flushleft}
\begin{tabular}{lr@{.}lcr@{.}l}
\hline\noalign{\smallskip}
 Element & \multicolumn{2}{c}{Values} & & \multicolumn{2}{c}{Errors} \\
\hline\noalign{\smallskip}
$P$ [d]   & 7&7570 & $\pm$ & &0005 \\
$T$ [HJD-2440000] & 8999&5 & $\pm$ & &3 \\
$e$       & 0&035 & $\pm$ & &009 \\
$\gamma$ [km s$^{-1}$] & 6&3 & $\pm$ & &4 \\
$\omega$ [$^{\circ}$] & 297& & $\pm$ & 13& \\
$K1$ [km s$^{-1}$] & 59&7 & $\pm$ & &6 \\
$K2$ [km s$^{-1}$] & 65&2 & $\pm$ & &8 \\
m$_1$ sin$^3i$ [M$_{\odot}$] & 0&82 & $\pm$ & &02 \\
m$_2$ sin$^3i$ [M$_{\odot}$] & 0&75 & $\pm$ & &02 \\
$a_1\sin i$ [Gm] & 6&36 & $\pm$ & &07 \\
$a_2\sin i$ [Gm] & 6&95 & $\pm$ & &08 \\
$\sigma (O-C)$ [km s$^{-1}$] & 2&39 & \multicolumn{2}{c}{} \\
$n_{obs}$ A & \multicolumn{2}{c}{25} & & \multicolumn{2}{c}{} \\
$n_{obs}$ B & \multicolumn{2}{c}{21} & & \multicolumn{2}{c}{} \\
\noalign{\smallskip}
\hline
\end{tabular}
\end{flushleft}
\end{table}

\subsection{Mean radial velocity}
The data contained in Table 1 form unique material to compute the mean
radial velocity of the Pleiades cluster. We have excluded from the list
all kind of binaries without an orbit and HII 2027 (variable systemic
velocity) and obtained a weighted mean value of 5.67 $\pm$ 0.56 (s.e.), based on 67 stars. We accounted for projection effects to compute these values: we used the convergent point of Rosvick et al. (1992a,b) to derive spatial velocities and we projected them on the direction perpendicular to the cluster center. The cluster mean radial velocity is defined in the IAU faint ($m_{v} \geq$ 4.3) standard system, as it has been discused by Mayor \& Maurice (1985). This standard system is slightly below the reference level defined by the minor planets, by about 0.4 km/s (Duquennoy et al. 1991).

\section{Binarity among the upper main sequence stars}
Radial velocities for the B- and A-type stars have been obtained by many 
authors, but no systematic study was conducted over a long time interval.
Morse et al. (1991) showed that it is possible to reach a precision of 1 
to 2 km s$^{-1}$ for B- and A-type stars by using a cross-correlation
technique. Liu et al. (1991), using a similar method, published a small
number of observations (1 to 3) for most member stars with a
spectral type earlier than F5.
 
\begin{figure}[t]
\centerline{\psfig{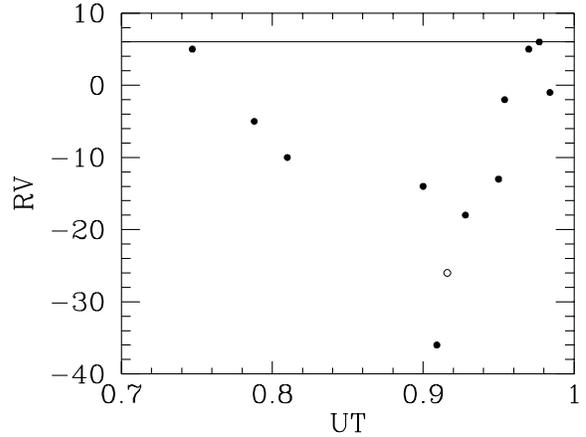}}
\caption[]{Radial-velocity observations on JD 2438302. The open circle
represents the $O-C$ residual for star HII 1397 from the orbit by
Conti (1968). This point fits well the observed trend during this night.}
\label{jd}
\end{figure}

The observations obtained at 128 {\AA}/mm by Abt et al. (1965), 
who made one of the major surveys in the Pleiades, suffer from systematic 
errors during some nights and several of the binaries detected have not 
been confirmed by other observers. To investigate the problem linked to Abt et al.'s (1965) observations, we have looked for stars that appear to be single in the ($V, U-B$) colour-magnitude diagram and constant in the radial velocity surveys by Smith \& Struve (1944) and Pearce \& Hill (1975). This defined a sample of 22 stars, and the nightly behaviour has been examined for several nights. Observation of subsamples of the 22 stars indicate that on some nights the average velocity of the observed stars (-10.1 to +15.5 km s$^{-1}$) is significantly different from that of the Pleiades mean velocity. Systematic trends have been noticed, as for example, on JD 2438302 (Fig.~\ref{jd}).

Consequently, the binary status of the early-type stars in the Pleiades
is not so well defined and more radial-velocity observations would be very 
useful. The poor present state of our binarity knowledge may appear discouraging when one thinks that the first radial velocity observations in the Pleiades were originally obtained in 1901-1908 with remarkable precision by Hartmann (Jung 1914, Hartmann 1914).

\section{The member list}
We have composed a large table collecting all 270 members brighter
than $V$ = 12.0 from Hertzsprung's (1947) catalogue and the corona 
samples discussed by Rosvick et al. (1992a) and Mermilliod et al. (1997). 
Table 3 contains the star identification, $V$ and $B-V$ from BDA,
the open cluster database (Mermilliod 1995), the $X$ and $Y$ rectangular
positions in arc minutes, the distances from the cluster center, the
multiplicity status, remarks and the deduced masses of stars and components. Table 3 is available only in electronic form from the Strasbourg anonymous ftp server (130.79.128.5).

The individual masses of the stars have been derived by different techniques
depending on the multiplicity status of the stars. For single stars, 
the mass has been computed from the $B-V$ colour with a power law relation between mass and $B-V$ derived from an isochrone calculated by the models of Schaller et al. (1992), for log t = 8.00 and z = 0.02.

For visual binaries, the colours have been computed from the magnitude
difference and checked by a photometric separation of the binaries
in the colour-magnitude diagram. From the distance above the ZAMS and
the local slope of the ZAMS, it is possible to compute reliable magnitudes
and colours of both components. The mass has then been computed from
the colour indices with the same power law as defined for the single stars. The same method has been used 
for the so-called photometric binaries, i.e. stars that are
clearly members, but without any positive binary detection from visual
observation, speckle interferometry or radial velocity, but still
located significantly above the ZAMS. For single-lined binaries, the
secondary mass has been taken within the interval defined by the 
spectroscopic minimum mass (when an orbit was available) and the maximum
mass defined by the fact that the star is 5 magnitudes fainter than
the primary. This condition is adopted because a difference of 5 magnitudes
between the primary and secondary produces an effect of 0.01 mag on the
observed magnitude of the system relative to the primary initial
magnitude. This procedure resulted in a probable excess of secondary masses
around 0.5 solar masses. The use of red colours, like $V-I$, would permit 
the detection of additional secondaries, because the secondary contribution
to the total flux is more important in the $I$ band than in $B$, and improve the mass determination. The values of the masses contained in Table 3 are those 
adopted to study the Pleiades structure and characteristics.

\section{Cluster structure}

\subsection{Global overview}

\subsubsection{Cluster flattening}

\begin{figure}[t]
\centerline{\psfig{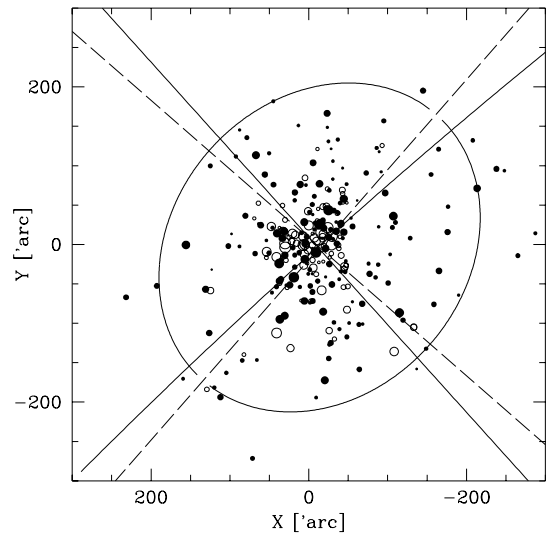}}
\caption[]{Map of the Pleiades displaying all the member stars considered in
this study. Open circles are multiple stars and filled circles stand for single stars. The point sizes are related to the star magnitudes. The solid lines represent the galactic longitude and latitude, and the dashed lines are the principal axes of the star distribution. The drawn ellipse encompasses 95 \% of the stars. North is at the top and East at the left of the map.}
\label{carte1}
\end{figure}

From the data collected in Table 3, we have plotted a chart of the Pleiades
(Fig.~\ref{carte1}) which displays all the member stars considered in the present study. The filled and open circles represent single and multiple stars respectively. The size of the circles is proportional to the star magnitudes.

\begin{figure}[t]
\centerline{\psfig{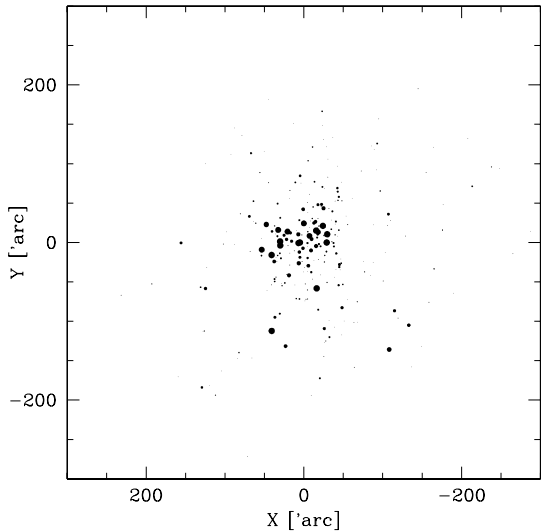}}
\caption[]{Map of the Pleiades displaying all the considered member stars . 
The point sizes are related to the masses of the stars or systems.}
\label{carte2}
\end{figure}

The difference between single and multiple star distribution is already apparent in Fig.~\ref{carte1}. The binary stars are more numerous in the central
part (R $<$ 80$\arcmin$). Does this feature depend on the binary character, i.e. on the greater cross-section of such systems, which favours the interactions with other stars, or only on the total mass of the system ? This point will be addressed in section 5.2 but Fig.~\ref{carte2} offers a first answer. It represents the same chart, but with the point size related to the stellar masses instead of the magnitudes. It is evident that the star distribution mainly depends on the star or stellar system masses.

Examination of density contour maps suggested that the cluster outer region is
elliptical in form. We applied a multi-component analysis to the data to derive the directions, and the ratio, of the ellipse axes. We found that the cluster halo is flattened roughly parallel to the galactic equator, with an ellipticity $e=(1-\frac{b}{a})$ of 0.17 (Fig. 3). We think that this effect is real and is not due to observational bias, because all stars listed by Artyukhina \& Kalinina (1970) as candidates from their proper motion survey, which extends out to 5$\degr$ from the Pleiades center, have been included in the CORAVEL observing list. 

Such a flattening of the outer part of open clusters is predicted both by theory and numerical simulations (Terlevich 1987). Furthermore, Wielen (1974) predicted that the ratios of the three orthogonal axes of the cluster, considered 
as a tridimensional ellipsoid, should be 2.0:1.4:1.0. The larger axis is pointing towards the galactic center and the smaller one is perpendicular to the galactic disk. As the Pleiades lie at a galactic longitude of 166$\fdg$6 and are close to us, we only observe the ratio of the second and third axes, namely 1.4:1.0, corresponding to an ellipticity of 0.29. However, 
the Pleiades are $\sim$ 50 pc below the galactic plane and we observe them under
an angle of 23$\fdg$5. We thus observe an effective axial ratio of 1.15:1.0 only, which corresponds to an ellipticity of 0.13. It is slightly less than the value we determined. It is worth noting that van Leeuwen (1983) already found a marginal flattening in the Pleiades, along the galactic plane. Oort (1979) found also a flattening in the Hyades, outside a 4 pc radius in the expected orientation, but the value is significantly greater than that predicted by theory. Finally, Gieseking (1981) observed a flattening roughly orthogonal to the galactic plane for NGC 3532, which is not predicted by theory.

Our results on the Pleiades confirm the theoretical expectations and van Leeuwen's results.

\subsubsection{Apparent star devoided area}
Another striking feature appearing on the maps shown in Fig.~\ref{carte1} and~\ref{carte2} is the apparent lack of stars at the western borderline of the Hertzsprung area,
at X $\sim$ -50$\arcmin$. This peculiarity is already noticeable in the maps presented by van Leeuwen (1983). If we plot cluster maps for stars with magnitudes in the intervals $V < 6$, $6 < V < 8$, $8 < V < 10$ and $10 < V < 12.5$, this feature is already present for $8 < V < 10$ and is very well defined for the last magnitude interval. The first likely explanation available is an observational bias due to the extension of the well observed Hertzsprung area, which ends just before this western discontinuity. Nevertheless, it is surprising that the same effect is \textit{not} so clearly observed on the three other sides of the Hertzsprung area. A more speculative explanation could be suggested by the work of White \& Bally (1993). Following their conclusions, the Pleiades are undergoing a supersonic encounter with an interstellar cloud, approaching the cluster from the \textit{west}. Thus it could be possible that the west part of the cluster suffers from a greater amount of extinction and therefore that this area appears less populated. However, the inspection of photometric data from the proper motion survey of Schilbach et al. (1995) does not allow us to confirm this prediction: the colours of the faint stars (down to magnitude $V$ = 18) located within the star devoided area are not different from those in other parts of the cluster. 

\subsection{Mass segregation}

\begin{figure}[t]
\centerline{\psfig{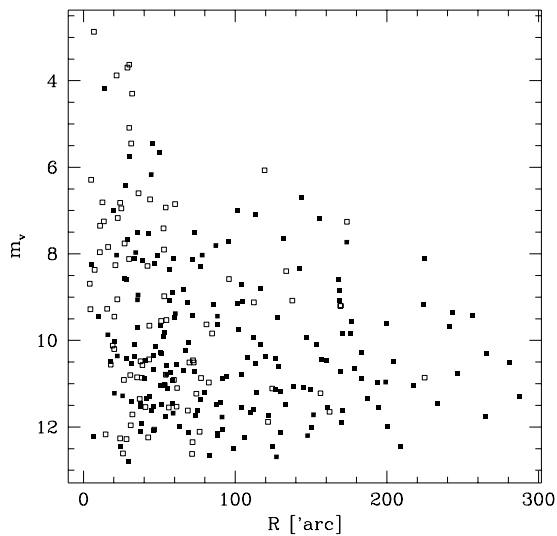}}
\caption[]{Apparent magnitudes of the stars as a function of their radial distances to the cluster center. The open and filled squares denote multiple and single stars
respectively.}
\label{VvsR}
\end{figure}

\begin{figure}[t]
\centerline{\psfig{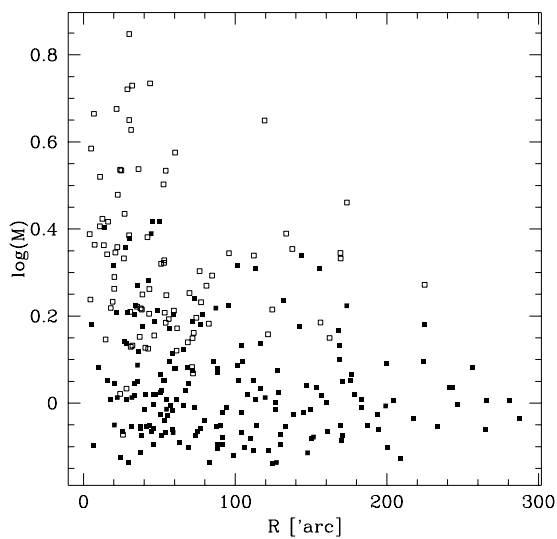}}
\caption[]{Logarithm of the star masses as a function of their radial distances to the cluster center. Same symbols as in Fig. 5.}
\label{lMvsR}
\end{figure}

The concentration of multiple stars relative to single ones, of bright stars relative to fainter ones, and of massive stars relative to less massive ones is apparent in Figs~\ref{carte1} and~\ref{carte2}. The segregation is even
more evident in Figs.~\ref{VvsR} and~\ref{lMvsR}, which show a very clear increase of the star minimum magnitudes, or a decrease of the maximum masses respectively, with increasing radius. The stars or systems more massive than 2.5 M$_\odot$ are all, except one, contained within a radius of 60$\arcmin$, while stars or systems with masses smaller by a factor of two are within a radius of 180$\arcmin$, i.e. a factor 3. These diagrams provide more information on the radial distribution of stars within a cluster than the usual radial distribution represented in the form of a histogram of the number stars at various distances from the cluster center. In particular they show a nice relation between the star luminosities or masses and their distances from the center which could help to estimate the cluster radius at a given stellar luminosity or mass. Figures~\ref{VvsR} and~\ref{lMvsR} also characterize the completeness of our sample in terms of magnitude, mass and radial extension. A similar figure, $V$ magnitude versus radial distance, made by combining the various recent surveys which extends to much fainter magnitudes (Stauffer 1996, Fig. 1) shows the distribution of members and candidates in the Pleiades. Large gaps are apparent and our knowledge of the Pleiades remains incomplete below K0 in terms of both magnitudes and area surveyed.

To investigate more accurately the radial distribution of the different star populations we have split the sample into two classes, according to different criteria: magnitude, mass and multiplicity of the stars.

\subsubsection{Single and multiple stars}

\begin{figure}[t]
\centerline{\psfig{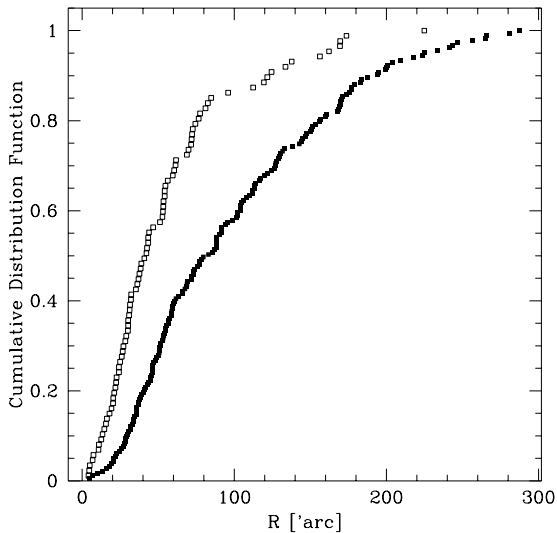}}
\caption[]{Cumulative distributions for the multiple stars (open squares) and the single stars (filled squares).}
\label{cumul3}
\end{figure}

In contrast to the mass/magnitude parameters for which it is difficult to find an objective value to divide the sample, the binary parameter is more direct: either the star is single or it is double or multiple and we can compare, without ambiguity, the cumulative distributions of the two populations (Fig.~\ref{cumul3}). The result is quite impressive and a Kolmogorov-Smirnov test clearly indicates that the two distributions are different. However, we should keep in mind that the concentration of multiple stars towards the cluster center may be overestimated. Indeed, the detection of low luminosity double or
multiple systems is probably more complete in the central area of the cluster.
In the central region we obtained more radial velocities over a longer period,
which increases the completeness of the binary detection, while, in addition,
near-infrared imaging has been conducted only in the Hertzsprung (1947) area by
Bouvier et al. (1997).

This would lead one to underestimate the number of such systems in the cluster outer part, and to slightly overestimate their concentration towards the cluster center. However, most stars resolved by Bouvier et al. (1997) were classified "PHB" by Mermilliod et al. (1992), i.e. they were identified as binaries from their position in the colour-magnitude diagram. Therefore, the same reasoning applied to the corona ensures that we missed a minimum number of binary stars.

Among the multiple star population itself, a mass segregation is obvious (Fig.~\ref{lMvsR}) and the distribution of multiple stars with masses greater than 2.5 M$_\odot$ (23 stars) is significantly different from the distribution of multiple stars with masses between 1.4 and 2.5 M$_\odot$ (52 stars). Moreover, if we divide the multiple star population between ``long period'' binaries (visual, IR imaging, speckle and occultation binaries; 44 systems) and ``short period'' binaries (spectroscopic binaries, 43 systems) we notice that their radial distributions are very similar. These results are consistent with the hypothesis that the radial segregation towards the cluster center depends mainly on the mass of the systems, and not on their periods. The same conclusion was obtained by Raboud \& Mermilliod (1994), and some theoretical considerations may help to understand these results.

Mathieu (1985) estimated that the period boundary between ``hard'' binaries, which are systems with a binding energy much greater than the mean kinetic energy of the stars in the cluster, and ``soft'' binaries lies at $10^{3}$-$10^{4}$ yr in the Pleiades, if the two binary components have a mass of 1 M$_\odot$. Such a period implies an apparent separation of 1$\farcs$ - 4$\farcs$7 and, as we have only 9 stars displaying this characteristic, we could consider that our sample is mainly composed of hard binaries (78 systems among 87). Following Hills \& Fullerton (1980) the hard binaries increase their binding energies, during close encounters with single stars, at the same average rate independently of their semi-major axes. Thus one cannot hope to detect any concentration among the binaries in function of the period. But we will observe a mass segregation between the binaries, resulting from energy equipartition, because the mass of the multiple systems are different.
However, Spitzer \& Mathieu (1980) show that during close encounters between two hard binaries the softer is disrupted. This scenario is only valid if one of the binaries is much harder than the other. In that case we should observe a greater proportion of hard binaries in the center of the cluster, where close binary encounters are most probable. Abt (1980) shows some evidence of such an effect, but we cannot confirm his result: our radial distributions of ``long'' and ``short'' period binaries cannot be considered as different.

\subsubsection{Sample subdivisions using a magnitude or a mass criteria}

Analysing the radial distribution of star populations defined by a magnitude or mass criteria implies the selection of a cut-off value. We could define this value as the one which optimises the separation between two well represented populations. However, such a selection is fuzzy and is very difficult to apply consistently to different clusters. Therefore, we decided to analyse the morphological modifications of the cumulative distribution functions for various cut-off to determine the value at which the two distributions become statistically different (between $m_{v}=4.0$ and 12.0, with a step of 0.1 mag, and between masses of 0.73 M$_\odot$ and 5.0 M$_\odot$, with a step of 0.05 M$_\odot$). Two distributions are considered different if the probability of false rejection of the null hypothesis (that the two distributions are the same) is less than 5\%, according to the test of  Kolmogorov-Smirnov.

\begin{figure*}[t]
\centerline{\psfig{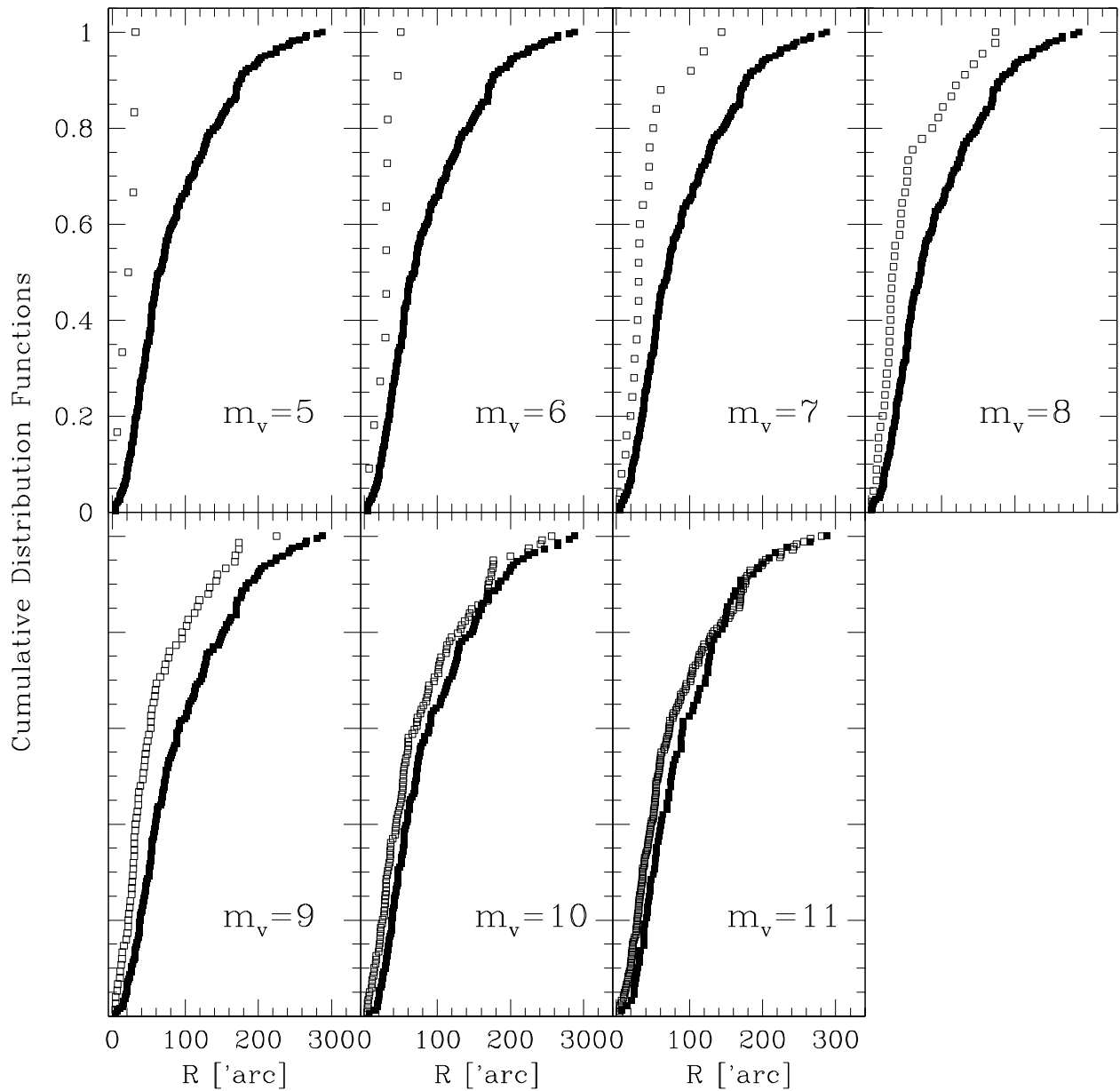}}
\caption[]{Cumulatives distributions for the bright stars (open squares) and the faint stars (filled squares). Cut-off values are indicated for the sample subdivisions.}
\label{cumulV}
\end{figure*}

Figure~\ref{cumulV} represents seven of the 81 sample subdivisions considered, for the magnitude criteria. These seven diagrams are sufficient to reveal the expected continuous behaviour of the two cumulative distributions for the various cut-off values. Interestingly, the two cumulative functions are considered as significantly different already for a cut-off value as great as $m_{v} = 10.9$, corresponding to masses M $\leq$ 1 M$_\odot$.

The same procedure has been applied to the sample divided according to the mass criterion. We found that the two cumulative distributions are significantly different for all cut-off values $M > 0.8$ M$_\odot$. This limit is smaller than the previous one. This difference results from the fact that, due to the high percentage of multiple stars in the central part of the cluster, the division of the sample in terms of magnitude underestimates the real mass difference between the two populations. 

The adopted procedure reveals that the definition of a universal cut-off value between \textit{two} populations based on the statistical signifiance of the difference between the distributions is hardly applicable, because the populations appear to be different even for a mass limit as low as 0.8 M$_\odot$.

\begin{figure}[t]
\centerline{\psfig{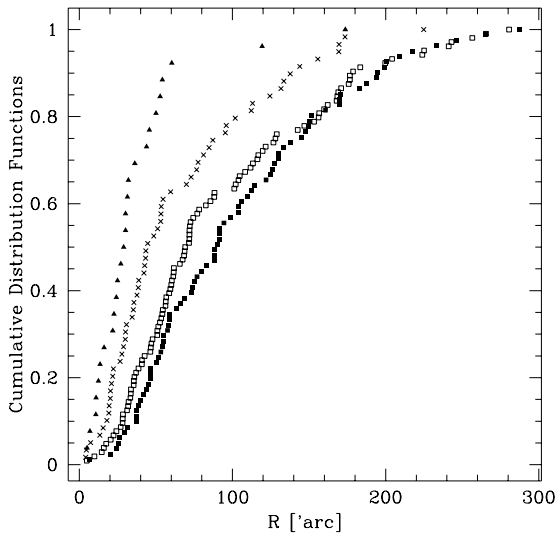}}
\caption[]{Cumulative distribution functions for different mass intervals: $M < 1$ M$_\odot$ (filled squares), $1 < M < 1.6$ M$_\odot$ (open squares), $1.6 < M < 2.5$ M$_\odot$ (crosses) and $M > 2.5$ M$_\odot$ (filled triangles).}
\label{cumulM}
\end{figure}

To characterize the degree of mass segregation among different populations in the cluster, it was better to subdivise the sample in more than two groups. 
Fig.~\ref{cumulM} represents the cumulative distributions for 4 different mass intervals. The Kolmogorov-Smirnov test indicates that the three more massive populations are significantly different, but that the distribution of the $M < 1$ M$_\odot$ and $1 < M < 1.6$ M$_\odot$ populations could not be considered as different. Even the radial distribution of the stars with masses $M < 0.8$ M$_\odot$ could not be considered as different from the population of stars with $1 < M < 1.6$ M$_\odot$. Consequently, mass segregation is less important for the lowest mass stars of the cluster, a result already noticed by van Leeuwen (1983), and also found in the Hyades (quoted by Mathieu 1985). It could be a general phenomenon because lower-mass star population is the most spatially extended, due to energy equipartition, and therefore is the most severely truncated by the galactic tidal field (Mathieu 1985).

\subsection{Characteristic radii}
A way to quantify the different radial distributions is to compute the values of characteristic radii, such as the radius containing half of the total number of stars ($r_{n/2}$), the radius containing half of the total mass of the stars ($r_{m/2}$), the core radius ($r_{c}$), the tidal radius ($r_{t}$) or the harmonic radius ($\overline{r}$).
The description of an apparent star distribution with characteristic radii supposes that the cluster has a spherical symmetry, which is not the case of the Pleiades, at least for the outer part. However, the study of a slightly flattened system is well achieved with the assumption of spherical symmetry and does not need the use of more sophisticated distributions. A simple simulation shows that the derivation of the core and tidal radii are unaffected by an artificial flattening of a cluster defined by a stellar density law of King (1962), with an axial ratio of 1.4:1 in order to reproduce the theoretically predicted ellipticity.

\subsubsection{Core and tidal radii}
The core and tidal radius were obtained by fitting, through the observed stellar-density distribution, the empirical density law of King (1962), given by
\begin{equation}
f(r)=k\left\{\frac{1}{\left[1+\left(\frac{r}{r_{c}}\right)^{2}\right]^{\frac{1}{2}}}-\frac{1}{\left[1+\left(\frac{r_{t}}{r_{c}}\right)^{2}\right]^{\frac{1}{2}}}\right\}^{2}
\label{eq:king}
\end{equation}
where $f(r)$ is the projected number density as a function of radius $r$, $k$ is a constant of proportionality, $r_{c}$ and $r_{t}$ are the core and tidal radii respectively. The adopted procedure for adjusting this density law could be divided in two parts: the determination of the initial values of $k$, $r_{c}$ and $r_{t}$ for the fitting process and the fitting process itself.

For the estimation of the initial parameters we computed a great number of King's stellar density laws, for the cluster divided in concentric rings, with $k$, $r_{c}$ and $r_{t}$ chosen by a Monte-Carlo procedure within large ``windows''. We then selected the set of parameters for which the function $\chi^{2}$ was minimum:
\begin{equation}
\chi^{2}=\sum_{i=1}^{n}\frac{1}{\sigma(i)}\left[dens_{obs}(i)-dens_{model}(i)\right]^{2}
\end{equation}
where $dens_{obs}(i)$ is the observed projected number density in the $i^{\mbox{th}}$ ring, $dens_{model}(i)$ is the King's law prediction of this density and $\sigma(i)$ is the uncertainty from Poisson statistics ($\sigma(i)=\frac{\sqrt{N(i)}}{A(i)}$, where $N(i)$ is the star number in the $i^{\mbox{th}}$ ring and $A(i)$ is the area of this ring).

This set of parameters was used as the initial one for the second part of the adjusting procedure: the fitting process itself. We estimated the best-fit parameters by minimizing the same $\chi^{2}$ function, in the standard manner. Incorrect distributions were rejected according to the criterion defined by Lampton et al. (1976). The estimated parameter uncertainties correspond to 95\% independent confidence intervals. These ranges were estimated using the procedure of Lampton et al. (1976). We thus obtained fitted values for $k$, $r_{c}$ and $r_{t}$.

The whole procedure is applied to 16 cluster subdivisions in 4 to 20 concentric rings. The final values of the density law parameters, reported in table~\ref{table_radius}, are the weighted mean of these 16 sets of values.

The tidal radius ought to be identical for all populations and independent of the star masses of their members. However, due to the difficulty to extrapolate the data towards the external parts of the cluster for the concentrated populations,
the values of $r_{t}$ given in Table~\ref{table_radius} are different. The only reliable estimations of the tidal radius are obviously those based on the more extended populations.

\subsubsection{Harmonic radius}
The harmonic radius $\overline{r}$ is defined following the equation of the potential energy of the cluster $\Omega$ (Chandrasekhar 1942):
\begin{equation}
\Omega=-G\sum_{i=1}^{n}\sum_{j>i}^{n}\frac{m_{i}m_{j}}{r_{ij}} \simeq -\frac{1}{2}\frac{GM^{2}}{\overline{r}}
\end{equation}
where $G$ is the gravitational constant, $m_{i}$ is the stellar mass, $r_{ij}$ is the distance between star $i$ and $j$ and $M$ is the total mass of the cluster.

If we estimate the apparent stellar density, $F(r)$, in strips of equal area covering the cluster, it is possible to relate the harmonic radius to this density through the relation (Schwarzschild 1954)
\begin{equation}
\overline{r}=\frac{(\int_{0}^{\infty}F(r)dr)^{2}}{\int_{0}^{\infty}F^{2}(r)dr}
\end{equation} 
In our study we have computed the apparent stellar density in 10 and 30 strips. Each density is smoothed by computing its average between 20 equidistant strips, with different orientations, uniformly distributed around the cluster center.
The uncertainty associated to the harmonic radius corresponds to $1\sigma$. It should be multiplied by a factor 2 to be compared with the uncertainty of $r_{c}$ and $r_{t}$ displayed in Table 4.

\subsubsection{Comparisons of the various radii}
Table~\ref{table_radius} presents the results for the radius
containing half of the total number of stars ($r_{n/2}$), the radius containing half of the total mass of the stars ($r_{m/2}$), the core ($r_{c}$), the tidal ($r_{t}$) and the harmonic ($\overline{r}$) radii for different distributions. As stated above, the subdivision criterion chosen for the bright/faint stars and the massive/less massive stars is not based on a statistical test. We adopt a cut-off value of $V$ = 9.5 because it corresponds to the limit between the A-F and the G-K stars. At spectral types F0-F5 occurs the 
transition from radiative to convective atmosphere which produces a gap
(B\"ohm-Vitense \& Canterna 1974) best seen in colour-colour diagrams. Due
to the position of this gap in the Pleiades main sequence, it is a good
limit to split the sample. The corresponding cut-off in terms of mass is 
$M$ = 1.5 M$_\odot$. It is worth noting that the two sub-samples created with the mass subdivision are not exactly the same as the ones created with the magnitude cut-off, because the mass subdivision takes more efficiently into account the multiple status of the systems present in the cluster.

\begin{table}
\setcounter{table}{3}
\caption{Characteristic radii [\arcmin] for different populations. The errors associated with $r_{n/2}$ and $r_{m/2}$ are typically between 5 and 10 [\arcmin].}
\label{table_radius}
\begin{flushleft}
\begin{center}
\begin{tabular}{|l c c c c r|}
\hline \rule{0pt}{1.2em}
{Population} & {$r_{n/2}$} & {$r_{m/2}$} & {$r_{c}$} & 
\multicolumn{1}{c}{$r_{t}$} & \multicolumn{1}{c|}{$\overline{r}$} \\
\hline
\textbf{Complete sample}     & 62 & 53 & 38 (14) & 445 (182) & 109 (32) \\
Bright stars        & 49 & 36 & 33 (25) & 241 (189) &  86 (24) \\
Faint stars         & 72 & 69 & 51 (25) & 434 (220) & 123 (28) \\
Massive stars       & 43 & 37 & 22 (14) & 327 (273) &  75 (21) \\
Less massive stars  & 77 & 73 & 58 (29) & 436 (224) & 134 (31) \\
Single stars        & 81 & 69 & 55 (27) & 461 (253) & 133 (29) \\
Multiple stars      & 42 & 35 & 32 (20) & 176 ( 79) &  70 (21) \\
\hline
\end{tabular}
\end{center}
\end{flushleft}
\end{table}

As expected, the different radii in Table~\ref{table_radius} behave similarly. The bright, massive and multiple populations are charaterised by a small radius, while the faint, less massive and single populations are characterised by a larger radius. Moreover, for each considered population the half-mass radius is always smaller than the radius containing the half total number of stars. This result means that we observe mass segregation even in each sub-sample of cluster stars. This again indicates that the segregation among cluster stars is mainly governed by the mass.

\subsection{The frequency of multiple star systems}

As the multiple stars are preferentially concentrated towards the cluster center, relative to single stars, it is interesting to compare their proportions in different parts of the cluster with predictions from numerical models.

Kroupa (1995) predicted the existence of a large proportion of binaries in the central region of star clusters, computed with the relation:
$$f=N_{binaries}/[N_{binaries}+N_{single\, stars}]$$
In particular, he expected that the total proportion of binaries in the central 2-pc sphere of the Pleiades is ``probably'' close to 60 \%. This
value is larger than that we obtained (48 \%) in the central 2-pc disk. However,
as is usually mentioned in the discussion of binary detection, the observed
value is most probably a lower limit. In the outer part of the cluster, the
binary frequency is 20 \% only. 

\begin{figure}[t]
\centerline{\psfig{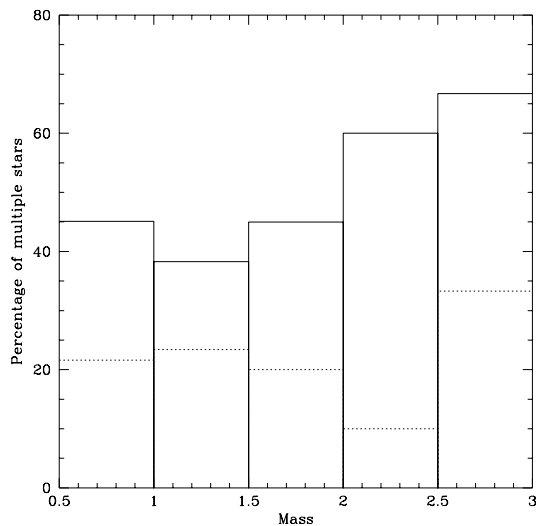}}
\caption[]{Percentage of multiple stars as a function of the primary mass, in the cluster central part. From the left to the right the bins involve: 28/23 (single/multiple stars); 29/18; 11/9; 4/6 and 3/6. The dotted lines correspond to the percentage of spectroscopic binaries, relative to the total number of stars.}
\label{permul}
\end{figure}

A binary fraction dependence on \textit{primary} mass is considered as a discriminator betweeen the ``capture'' binary formation mode and the fragmentation mode. In the first model the binary fraction is a strongly increasing function of primary mass, but in the second one the binary fraction is only weakly dependent on primary mass, and then in the sense of binary fraction declining with increasing primary mass (Clarke 1997). As the detection of spectroscopic binaries in the halo of the Pleiades is probably not complete, we restricted our study of the percentage of multiple systems, as a function of the \textit{primary} masses, to the Hertzsprung area. Figure~\ref{permul} represents the values observed among this sub-sample, indicating a possible dependence of the multiple star fraction with primary mass. Six among the 7 brightest cluster stars belong to multiple systems (Fig.~\ref{VvsR}), i.e. the most massive stars are multiple ones.

Before claiming this effect has a physical origin we need to examine possible biases due to statistics and detection limits. Statistically, if we consider a star cluster only composed of binaries, which were formed by \textit{random pairing}, we could expect a decrease of the probability to observe a star as a \textit{primary} with decreasing mass (Kroupa et al. 1996). In such a cluster, the most massive star has a probability of 100 \% to be observed as a \textit{primary}. However, considering a Salpeter mass function ($f(m)=Cm^{-(1+x)}$, with $(1+x)=2.35$) extending down to 0.08 M$_\odot$, we found that a 0.7 M$_\odot$-star (i.e. the least massive stars in our sample) has a probability of 95 \% to be observed as a \textit{primary}: obviously the mass range from which a secondary can be picked decreases with decreasing mass of the primary. Using the mass function exponent we derived for our sample ($(1+x)=2.5$, see next section), we obtain a probability of 96 \% to observe a 0.7 M$_\odot$-star as a primary. This probability is only 82 \% if we consider the exponent values used by Kroupa et al. (1991) and McDonald \& Clarke (1993) (2.35 for stars with mass $m > 1$ M$_\odot$, 1.1 for $0.08 \leq m \leq 0.5$ M$_\odot$ and 2.2 for $0.5 < m \leq 1$ M$_\odot$).

On the other hand, the detection of low mass secondaries in spectroscopic binaries is more difficult for B- and A-type stars. So, if any, the detection biases would lead to underestimate the binary frequency for star with mass larger than 2 M$_\odot$.

As our frequency of multiple systems range from $\sim$100\% for the brightest sample stars to around 40 \% for the less massive stars, we conclude that this effect could not be imputed to the mentioned statistical, or detection, biases. This conclusion also implies that the binaries are not created by random pairing, if the effects of dynamical evolution can be neglected.

The results presented in Fig.~\ref{permul}, and the fact that the most massive cluster stars belong to multiple systems, could be explained by the effect of dynamical evolution instead of a different formation mode. Encounters with exchange reactions, during which massive single stars replace lower mass binary components is a likely occurring phenomenon (Mathieu 1985). In a 100-body equal-mass cluster with an 11 \% population of slightly hard binaries, 24 crossing times are enough in order that approximately two thirds of the binaries had encounters (Aarseth 1975).

The Pleiades have an age of $10^{8}$ yr and therefore do not present a snapshot of pure star formation products. Thus our results cannot be used to discriminate between different binary formation scenarios. We need to repeat the same kind of investigations in very young (and populous) open clusters to be able to constrain theoretical models. However, it is worth noting the current lack of such a suitable observational sample, because sufficiently populous young clusters are distant objects suffering from extreme contamination problems.

\subsection{The mass functions}

The improved material collected in this paper allows various determinations of the Pleiades mass function, for the complete sample, for single stars and primaries only, and allows examination of the effect of radial extension.

We fitted a Salpeter-type power law in the form
\begin{equation}
log\left(\frac{df}{dM}\right)=C-(1+x)\;log(M)
\label{eq:fctm}
\end{equation}
throughout the observed data. In Eq.~(\ref{eq:fctm}) $df/dM$ is the number of stars per unit mass as a function of mass $M$. $C$ is a constant and $(1+x)$ is the power law exponent, which has the value of 2.35 following Salpeter (1955).

We first derived the mass function for the complete sample, without any correction for the binary content of the cluster. We therefore computed all the stellar masses with our relation between $(B-V)$ and the mass (Sect. 4). The mass function that we obtained is polluted with unresolved binaries (case 1 in Table~\ref{x}). The derived mass function slope (2.1 $\pm$ 0.21) agrees marginally with the canonical value of Salpeter (1955).

Using the available binarity information of our sample, we derived a mass function only for the single stars and the primaries for the complete sample (case 2 in Table~\ref{x}). The slope of 2.5 $\pm$ 0.15 is not very different from the previous one, indicating that the determination of mass function is not seriously affected by unresolved multiple stars. This finding confirms the results of Tarrab (1982).

\begin{figure}[t]
\centerline{\psfig{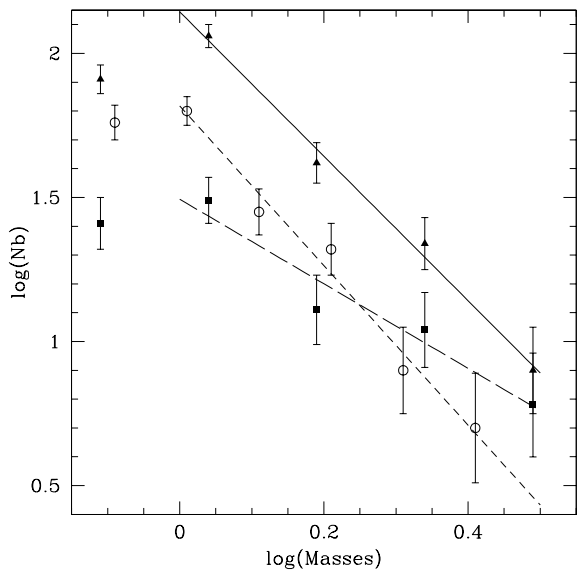}}
\caption[]{Mass functions. The solid line stands for the complete sample (single stars and primaries). The long-dashed line represents the mass function of the primaries and the short-dashed line stands for the single stars.}
\label{fmas}
\end{figure}

It is also very interesting to compare separately the mass functions of the single stars and of the primaries of multiple systems. Following Vanbeveren (1982), the similarity between these two kinds of function is rather improbable. He also found, in the case where the probability of the formation of binaries increases with increasing cloud mass, that the initial mass function of the primaries is less steep than for the single stars. These computations were done with masses between 15 M$_\odot$ and 150 M$_\odot$. Although our mass range is very different, it is worth noting that our results (cases 3 and 4 in Table~\ref{x} and Fig.~\ref{fmas}) agree with these conclusions. This result is expected from the observations presented in the previous section about the percentage of multiple systems as a function of the primary masses.

Different initial mass functions for single stars and primaries would imply that these two populations have different origins, and therefore discredit the ``capture'', in the sense of \textit{random pairing}, formation mode for the binaries. However, as noted before, we are not observing the initial conditions of stellar formation and the currently different mass functions of the single stars and the primaries may result from the dynamical evolution of the cluster. Due to encounters with exchange reactions, the low mass primaries would tend to be replaced by massive single stars, decreasing the slope of the primaries mass function relative to the single stars mass function as it is observed. If we now restrict the comparison of these two mass functions to the Hertzsprung area, where the multiple stars detection is more complete but where the mass segregation effect has depleted the low mass stars, we still observe the same results. The slope $(1+x)$ is 1.57 $\pm$ 0.34 for the single star mass function and 0.68 $\pm$ 0.16 for the primary mass function.

To test the effect of an incomplete surface coverage of the cluster field on the power law exponent, we only considered the stars within the Hertzsprung area (case 5 in Table~\ref{x}). The mass function slope is then less steep than for the complete sample (case 2). This result is evidently due to mass segregation, which depletes the inner cluster part from the less massive stars (Pandey et al. 1991a, b). It is therefore very important to consider the cluster and its surrounding halo in order to derive a realistic mass function, as already stated by Scalo (1986). If we only consider the outer part of the cluster (case 6) we obtain a very steep slope for the mass function, also expected from the effect of mass segregation, as this area is highly depleted in massive stars.

Finally, if we compute the mass function of the cluster by considering the stellar system masses merely as the sum of the different component masses (cases 7 and 8), we obtain smaller slopes for the complete cluster and for the Hertzsprung area than for the corresponding single and primaries mass functions (cases 2 and 5). This results because the mass summed method ``creates'' massive stars which have no physical reality. Nevertheless the effect of mass segregation is still noticeable.

\begin{table}
\caption{Values of different power law exponents $(1+x)$}
\label{table xx}
\begin{flushleft}
\begin{center}
\begin{tabular}{|l c|}
\hline \rule{0pt}{1.2em}
{Sample} & {$(1+x)$} \\
\hline
(1) Complete sample (with unresolved binaries)           & 2.10 $\pm$ 0.21 \\
(2) Complete sample (singles + primaries)       & 2.50 $\pm$ 0.15 \\
(3) Singles                   & 2.75 $\pm$ 0.22 \\
(4) Primaries                 & 1.47 $\pm$ 0.27 \\
(5) Hertzsprung sample (singles + primaries) & 1.71 $\pm$ 0.14 \\
(6) Cluster outer part (singles + primaries) & 3.20 $\pm$ 0.16 \\
(7) Complete sample (mass summed)           & 1.76 $\pm$ 0.15 \\
(8) Hertzsprung sample (mass summed)        & 1.22 $\pm$ 0.18 \\
\hline
\end{tabular}
\end{center}
\end{flushleft}
\label{x}
\end{table}

\subsection{Estimation of the cluster total mass}

The cluster total mass is an important parameter for the comparison between results from numerical simulation of open cluster dynamical evolutions and real objects (Wielen 1975). However, its derivation is very difficult (Bruch \& Sanders 1983) as will be shown in this section. We use three methods to estimate the cluster total mass: the tidal radius, the Virial theorem and the mass function.

\subsubsection{The tidal radius}

The total mass of an open cluster is related to the tidal radius through
\begin{equation}
M_{c}=\frac{4A(A-B)}{G}r_{t}^{3}
\label{eq:mc}
\end{equation}
where $G$ is the gravitational constant, $r_{t}$ is the tidal radius of the cluster, $A$ and $B$ are Oort's constants of galactic rotation. This equation is directly derived from King (1962), with the assumption that the cluster is at the same distance from the galactic center as the Sun.

The tidal radius considered in Eq.~(\ref{eq:mc}) is measured in the direction of the galactic center. However, we only observe the tidal radius perpendicular to this direction but parallel to the galactic disk. In Sect. 5.1.1 we found that the Pleiades have a flattening close to the expected one. We could therefore consider that we have a cluster with axes ratios 2:1.4:1 and then the tidal radius in the direction of the galactic center has a value of $2/1.4$ times the value of the observed tidal radius (445\arcmin\, from Table 4).

Using $A=15$ km s$^{-1}$, $B=-12$ km s$^{-1}$ and $r_{t}=636\arcmin = 23$ pc, we derive a total cluster mass of 4000 M$_\odot$. The confidence interval, based only on an uncertainty of 1$\sigma$ for $r_{t}$, is 1600 to 8000 M$_\odot$. If we do not apply any correction due to the cluster flattening we obtain 1400 M$_\odot$ for the mass, with a confidence interval between 530 and 2900 M$_\odot$. As may be seen from Eq.~(\ref{eq:mc}), this cluster mass determination is very sensitive to the tidal radius value, which is only poorly determined.

\subsubsection{The Virial theorem}

To derive the cluster mass through the Virial theorem, we need to estimate the velocity dispersion in the cluster. From 67 member stars, located in the Hertzsprung area (see section 2.3), we obtain a radial velocity dispersion of 0.36 km s$^{-1}$, from the quadratic difference of the total observed dispersion corrected for projection effects (0.56 km s$^{-1}$) and the mean measurement error (0.43 km s$^{-1}$). Assuming velocity isotropy and taking for the harmonic radius $\overline{r}=109$ ['arc] $= 3.96$ pc, we derive a cluster total mass of 720 M$_\odot$, through the relation
\begin{equation}
M_{c}=\frac{6\;\overline{r}\;\overline{V_{r}^{2}}}{G}.
\end{equation} 
The confidence interval, based only on the harmonic radius uncertainty, is 510 to 940 M$_\odot$,   

\subsubsection{The mass function}

Finally, we estimate the cluster mass using the observed mass function (see Sect. 5.5). We assume that the mass function for the single and primaries (case 3 in Table~\ref{x}) is applicable down to 0.08 M$_\odot$ and we include a factor 1.16 to take into account binary companions (32 \% of the stars are binaries with a companion mass equal to the half of the primary mass). The result is a cluster total mass of 950 M$_\odot$, with a confidence interval between 800 and 1150 M$_\odot$, based only on the mass function slope uncertainty.

\subsubsection{Pleiades total mass}

The three different mass estimations (Table~\ref{masse}) are relatively close together, within a factor 2, except the first determination taking into account a correction for the cluster flatness, which is a factor more than 5 greater than the lowest mass estimation. van Leeuwen (1983) estimated a cluster total mass of 2000 M$_\odot$, which is within our confidence intervals of mass determination through the tidal radius.

\begin{table}
\caption{Results of the different cluster total mass determinations (see text for more details).}
\label{table xx}
\begin{flushleft}
\begin{center}
\begin{tabular}{|l c c|}
\hline \rule{0pt}{1.2em}
{Method} & \multicolumn{1}{c}{Cluster total mass} & \multicolumn{1}{c|}{1 $\sigma$ confidence interval}\\
 & {M$_\odot$} & {M$_\odot$}\\
\hline
\textbf{Tidal radius}   & 4000 & [1600, 8000] \\
\multicolumn{3}{|l|}{(with correction for the cluster flatness)} \\
\textbf{Tidal radius}   & 1400 & [\, 530, 2900] \\
\multicolumn{3}{|l|}{(without correction for the cluster flatness)} \\
\textbf{Virial theorem} & \, 720 & [\, 510, \, 940] \\
\textbf{Mass function} & \, 950 & [\, 800, 1150]\\
\textbf{Summed mass} & \, 412 & \\
\hline
\end{tabular}
\end{center}
\end{flushleft}
\label{masse}
\end{table}

If we sum up all the stellar masses derived for our whole sample of stars, we obtain 412 M$_\odot$. Then the stars down to $m_{v}=12.5$ and extending out to 5\degr\, from the cluster center represent at best $\sim$ 57 \% and at worst $\sim$ 10 \%  of the total cluster mass.
Member stars fainter than $V$ = 12.5 are already known in the Pleiades and
about 600 flare stars (spectral type K2 and later) have been detected. Assuming a mean mass of 0.5 M$_\odot$ and 400 faint members, not all flare stars are members (Jones 1981), we can add at least 200 solar masses.

These results clearly point out the great difficulties of estimating the total cluster mass with precision. All the estimators used are based on strong hypothesis and are dependent on poorly determined parameters. The only way to properly compute a cluster mass is to sum all the masses of the individual cluster members. 

\section{Conclusion}

A study of the Pleiades structure has been performed on the basis
of the presently available data which limits the sample to stars brighter
than $V$ = 12.5. We used the best present knowledge on duplicity in the
Pleiades. 

Using a multi-component analysis applied to the apparent stellar positions we find an ellipticity in the cluster outer part, in agreement with theoretical expectations and van Leeuwen's results (1983).

We have observed a clear mass segregation, which depends on the mass of the 
stars or systems and not on the binary periods. Consequently, binaries are more concentrated than single stars and massive binaries are more concentrated than less massive ones. The mass segregation is significant down to 1 M$_{\odot}$. Different radii have been computed to characterize the radial distributions of various cluster star populations.

For the first time, to our knowledge, the mass function of single stars and primaries of multiple systems have been determined separately and compared. They turned out to be different, in agreement with predictions made by Vanbeveren (1982). Such a result, if confirmed by subsequent studies, may have important implications for star formation models. It shows the extreme interest of detailed studies of young, or very young, open clusters and, especially, of their binary populations.

We review the great difficulty in deriving an estimation of the cluster total mass. The mass estimates span an order of magnitude, from $\sim$500 to 8000 M$_\odot$, once we consider parameter 1$\sigma$ errors. Considering only the stars included in our sample we derived a projected spatial mass density of 17.7 M$_\odot$ pc$^{-2}$, or 9.5 stars pc$^{-2}$, for the 2-pc radius central disk of the cluster. This region was chosen because it is widely used by Kroupa (1995) for comparisons between numerical models. Central density values for different clusters should be compared only if they are computed in the same manner, for example inside this central 2-pc radius disk.

There seems to be no other alternative to determine the Pleiades total mass and complete luminosity function than identifying all members in an area even larger than that investigated here and to fainter magnitude. The deepest surveys made in the Pleiades have been limited to the central region within a radius of 3$\degr$. They should be extended to at least 6$\degr$ from the center.

\acknowledgements{We thank Dr J. Bouvier for the communication, before publication, of his data concerning the multiplicity of the Pleiades stars obtained with IR imaging. We are also very grateful to Dr C. Prosser, the referee, whose valuable comments improved the paper.}

\end{document}